\documentclass[11pt]{article}
\usepackage[a4paper, portrait, margin=2.5cm]{geometry}
\usepackage{graphicx}
\usepackage{lastpage}
\usepackage{fancyhdr}

\fancypagestyle{firstpage}{%
  \setlength\headheight{154pt}
  \fancyhf{}%
  \fancyhead[C]{\includegraphics[width=\textwidth]{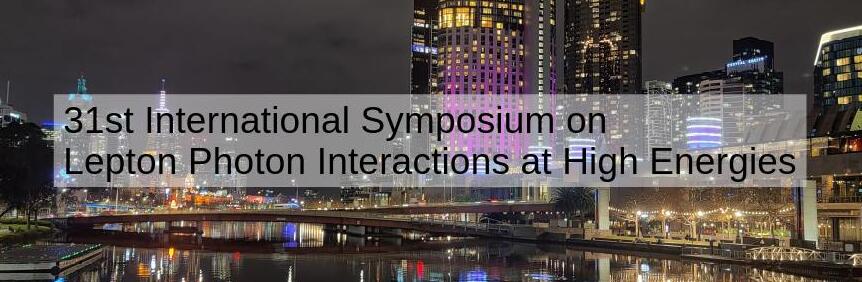}}%
  \fancyfoot[C]{CC-BY-4.0 licence}%
}

\begin{document}
\title{Dark sector and Axion-like particle search at BESIII} 
\date{17-21 July 2023}
\author{Vindhyawasini ~Prasad \\ Instituto de Alta Investigation \\
Universidad de Tarapaca \\
Casilla 7D, Arica, Chile \\ \textit{On behalf of the BESIII collaboration}} 

\newgeometry{top=2cm, bottom=7cm}
\maketitle
\thispagestyle{firstpage}
\abstract{Dark matter (DM) refers to a new type of matter that may explain observed rotation curves of galaxies and the composite structure of the Universe. It may couple to the Standard Model particles via portals, which include the possibility of axion-like particle, light Higgs boson, dark photon and spin-1/2 fermions. The axion-like particle and light Higgs boson can be accessible via radiative decays of $J/\psi$ while the dark photon via initial-state radiation process using the data of high-intensity $e^+e^-$ collider experiments, such as the BESIII experiment. DM may  be depicted as baryonic matter in an invisible final state. The presence of a massless dark photon, predicted by the spontaneous broken of Abelian group $U(1)_D$, may enhance the branching fractions of rare flavor changing neutral current decay processes. BESIII experiment has recently explored the possibility of these DM scenarios using the data samples collected at several energy points, including $J/\psi$ and $\psi(3686)$ resonances. This report summarizes the recent results of the BESIII experiment related to the dark-sector and Axion-like particles. 
}
\restoregeometry

\noindent
{\bf\em\boldmath Introduction:}
The Standard Model (SM) of particle physics is incredibly successful but unable to accommodate the features of recent experimental anomalies~\cite{dm,anoml}, such as dark matter (DM)~\cite{dm}. DM amounts around $27\%$ of total matter density of Universe. It neither emits nor absorbs electromagnetic radiation. So far, the presence of DM is inferred via gravitational effects on visible matter only. Thus, the nature of DM is still elusive.  A GeV scale DM carrying baryon number provide an attractive framework to understand the origin of dark matter and the matter-antimatter of the Universe~\cite{barioDM}.  The massless dark photon~\cite{masslessdm, tandean} may also enhance the decay rate of flavor changing neutral current (FCNC) processes, highly suppressed in the charm-sector~\cite{glashow}.  An experimental information of DM is necessary to explain observed rotation curves of galaxies and the creation of our Universe. Many extensions of the SM motivate a new type of \rm{\lq hidden-dark-sector\lq} under which the WIMP like dark matter is charged through a new type of dark force carrier~\cite{arkani}. The corresponding gauge field can couple to the SM particles via portals~\cite{Essig}, which could be a light Higgs boson, an axion like particle  (ALPs), a dark photon ($\gamma'$)  or spin-1/2 particles. If the masses of these particles are in the MeV/$c^2$ to GeV/$c^2$ range, they can be accessible by high-intensity $e^+e^-$ collider experiments, such as the BESIII experiment~\cite{bes3}. BESIII~\cite{bes3}, a symmetric $e^+e^-$ collider experiment running at  the  tau-charm region, has explored the possibilities of these DM scenerios using the data-sets collected at several enegy points, including at the  $J/\psi$ and $\psi(3686)$ resonances. This report reviews the recent results of the BESIII experiment related to the axion-like particle and dark matter searches. 

\noindent
{\bf\em\boldmath  Search for an axion-like particle:}
Axion is a pseudoscalar particle predicted by Peccei and Quinn~\cite{Peccei} to solve the strong $CP$~\cite{Weinberg} and hierarchy~\cite{Graham} problems in quantum chromodynamics. The ALPs have arbitrary masses and couplings. They are predicted by various extensions of the SM, such as extended Higgs sector~\cite{higgs} and string theory~\cite{string}.  At BESIII, the ALPs, $a$, can be accessible via ALP-Strahlung process $e^+e^- \to \gamma a$~\cite{Merlo} and radiative decay $J/\psi \to \gamma a$ process~\cite{Merlo, Masso}. ALP predominantly couples to a photon-pair with a coupling constant $g_{a \gamma \gamma}$. The experimental bounds on $g_{a \gamma \gamma}$ in the ALP mass region of $0.16 \le m_a \le 8$ GeV/$c^2$ are less constrained than  other $m_a$ regions. In this $m_a$ region, most stringent exclusion limits on $g_{a \gamma \gamma}$ mainly come from $e^+e^- \to \gamma \gamma(\gamma)$ process~\cite{opal}, ALP-Strahlung process $e^+e^- \to \gamma a$~\cite{belle2} and radiative decays of $J/\psi$ based on 2.7 billion of $\psi(2S)$ events at BESIII~\cite{bes3alp}. The exclusion limits on $g_{a \gamma \gamma}$ in the mass region of [0.16, 3.0] GeV/$c^2$ can be further improved by using 10 billion of $J/\psi$ data sample recently collected by the BESIII detector~\cite{nJps}.  

\begin{figure*}[htbp]
\centerline{\includegraphics[width=15.0cm]{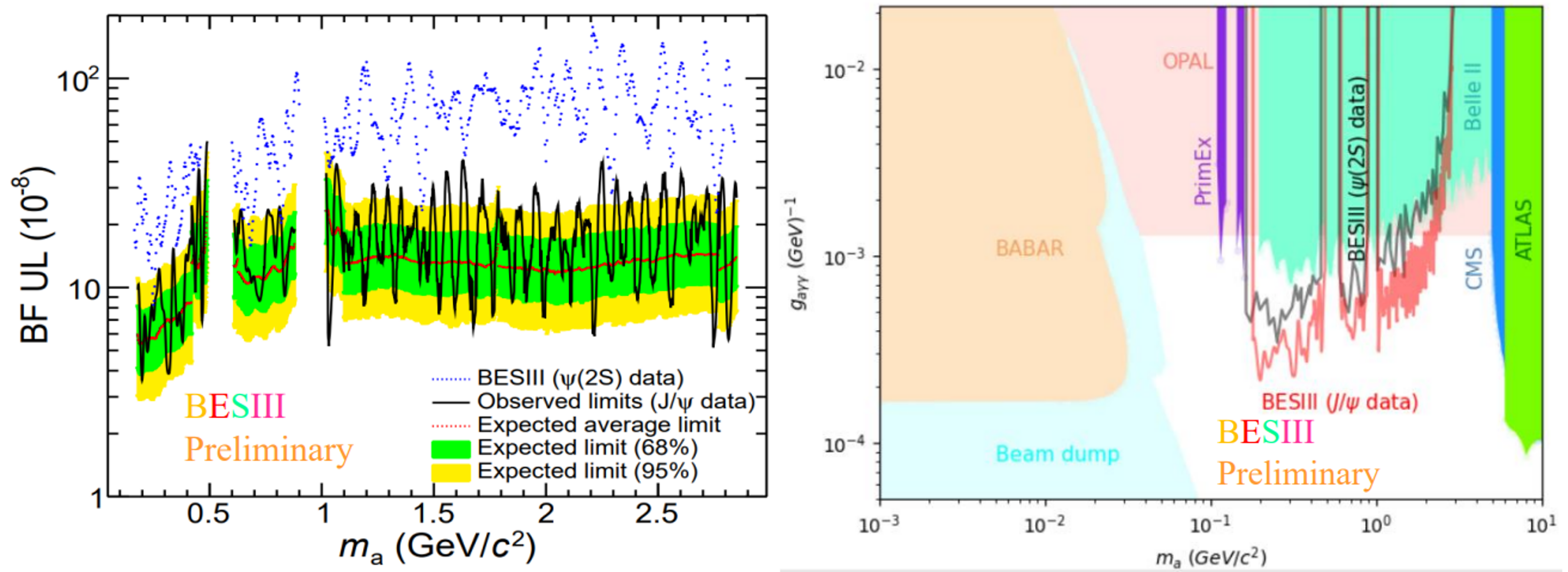}}
\caption{The $95\%$ CL upper limits on (left) product branching fractions $\mathcal{B}(J/\psi \to \gamma a) \times \mathcal{B}(a \to \gamma \gamma)$ together with the previous BESIII measurement and expected upper limit bands, and (right) ALP-photon coupling $g_{a \gamma \gamma}$  together with existing experimental bounds as a function of $m_a$.  All the results are preliminary. \label{mgg}}
\end{figure*}

BESIII has recently  searched for di-photon decays of ALP in radiative decays of $J/\psi$ using 10 billion of $J/\psi$ data. The event of interest is selected with at least three-photon candidates in the electromagnetic calorimeter (EMC) barrel region. The mass resolution is improved by performing a four constraint (4C) kinematic fit for $J/\psi \to \gamma \gamma \gamma$ reconstruction. The search for narrow ALP is performed in the steps of $1-2$ MeV/$c^2$ with a series of extended maximum likelihood fits to the di-photon invariant mass spectrum, $m_{\gamma \gamma}$, of all the three combinations of photons after rejecting the backgrounds from $J/\psi \to \gamma P$ ($P = \pi^0, \eta, \eta', \eta_c$) decays. No evidence of significant narrow ALP production is found in the $J/\psi$ data set at any ALP mass point. We set $95\%$ confidence level (CL) upper limit on the product branching fractions $\mathcal{B}(J/\psi \to \gamma a) \times \mathcal{B}(a \to \gamma \gamma)$ and ALP-photon coupling $g_{a \gamma \gamma}$ as a function of ALP mass, using the formula of Eq.(1) of Ref.~\cite{bes3alp}, in the ranges of $(3.6 - 53.1) \times 10^{-8}$ and $(2.2 - 97.5) \times 10^{-4}$, respectively, for $0.18 \le m_a \le 2.85$ GeV/$c^2$. The corresponding results are summarized in Fig.~\ref{mgg}. The new BESIII limits on $g_{a \gamma \gamma}$ are more restrictive than the existing ones~\cite{opal, bes3alp, belle2} and have an improvement by a factor of $2-3$ over the previous BESIII measurement~\cite{bes3alp} in the search mass region.

\begin{figure*}[htbp]
\centerline{\includegraphics[width=15.0cm]{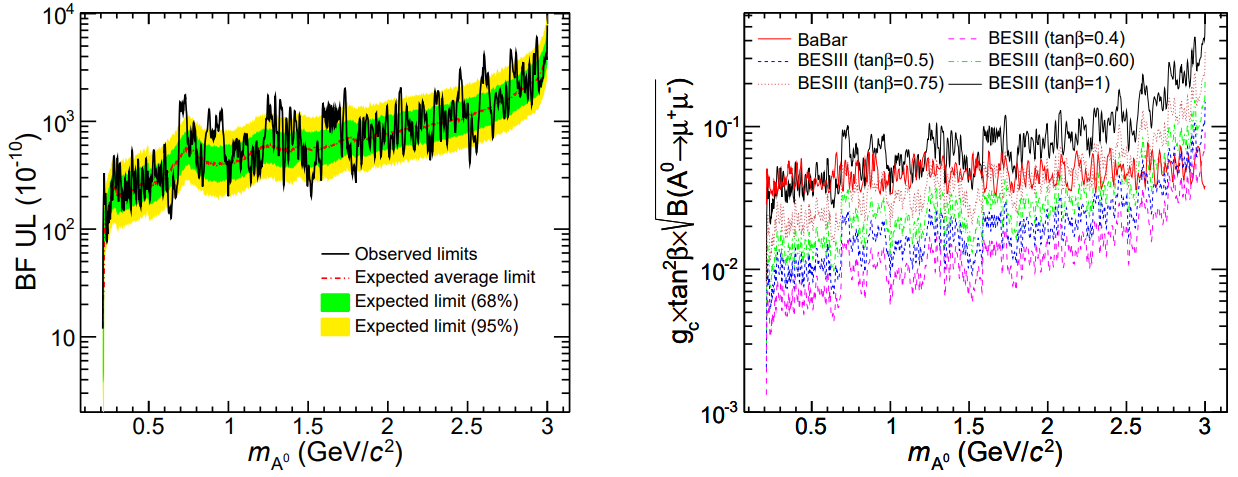}}

\caption{The $90\%$ CL upper limits on (left)  product branching fractions $\mathcal{B}(J/\psi \to \gamma A^0) \times \mathcal{B}(A^0 \to \mu^+\mu^-)$ together with expected ones and (right) effective Yukawa coupling of Higgs field to the bottom quark-pair for different values of $\tan \beta$ together with BaBar measurement as a function of $m_{A^0}$.  \label{higgs}}
\end{figure*}

\noindent
{\bf\em\boldmath  Search for \boldmath{$CP$}-odd light Higgs boson:}
 Many supersymmetric extensions of the SM, such as Next-to-Minimal Supersymmetric Model (NMSSM)~\cite{Maniatis}, predict the light Higgs boson. The Higgs-sector of the NMSSM contains seven Higgs bosons, among them, there is a $CP$-odd light Higgs boson ($A^0$) whose mass is expected to be a few GeV/$c^2$. The effecitive Yukawa coupling of the up (down) type of quark-pair is proportional to $\cot\beta$ ($\tan \beta$), where $\tan \beta$ is a standard SUSY parameter. The $A^0$ can be accessible via radiative decays of $V~(=J/\psi, \Upsilon(1S))$ mesons~\cite{Wilczek}. The branching fraction of $J/\psi \to \gamma A^0$ is expected to be within range of $10^{-9} - 10^{-7}$, depending upon $m_{A^0}$, $\tan\beta$ and NMSSM parameters~\cite{Dermisek}. The search for the $A^0$ have been performed with its various decay channels by many collider experiments, including BaBar~\cite{baber} and BESIII~\cite{prevbes3} experiments. But so far, only negative results are reported. BESIII has recently searched for di-muon decays of the $A^0$ in radiative decays of $J/\psi$ using 9 billion of $J/\psi$ events~\cite{bes3higgs}. No evidence of $A^0$ production is found and $90\%$ CL upper limits are set  on the branching fraction products $\mathcal{B}(J/\psi \to \gamma A^0)\times \mathcal{B}(A^0 \to \mu^+\mu^-)$ in the range of $(1.2 -778.0) \times 10^{-9}$ for $0.212 \le m_{A^0} \le 3.0$ GeV/$c^2$. The new measurement has a $6-7$ times improvement over the previous BESIII measurement~\cite{prevbes3} (Fig.~\ref{higgs} (left)). The new BESIII result is also slightly better than the BaBar measurement~\cite{baber} in the low-mass region for $\tan \beta =1$ (Fig.~\ref{higgs} (right)).

\begin{figure*}[htbp]
\centerline{\includegraphics[width=15.0cm]{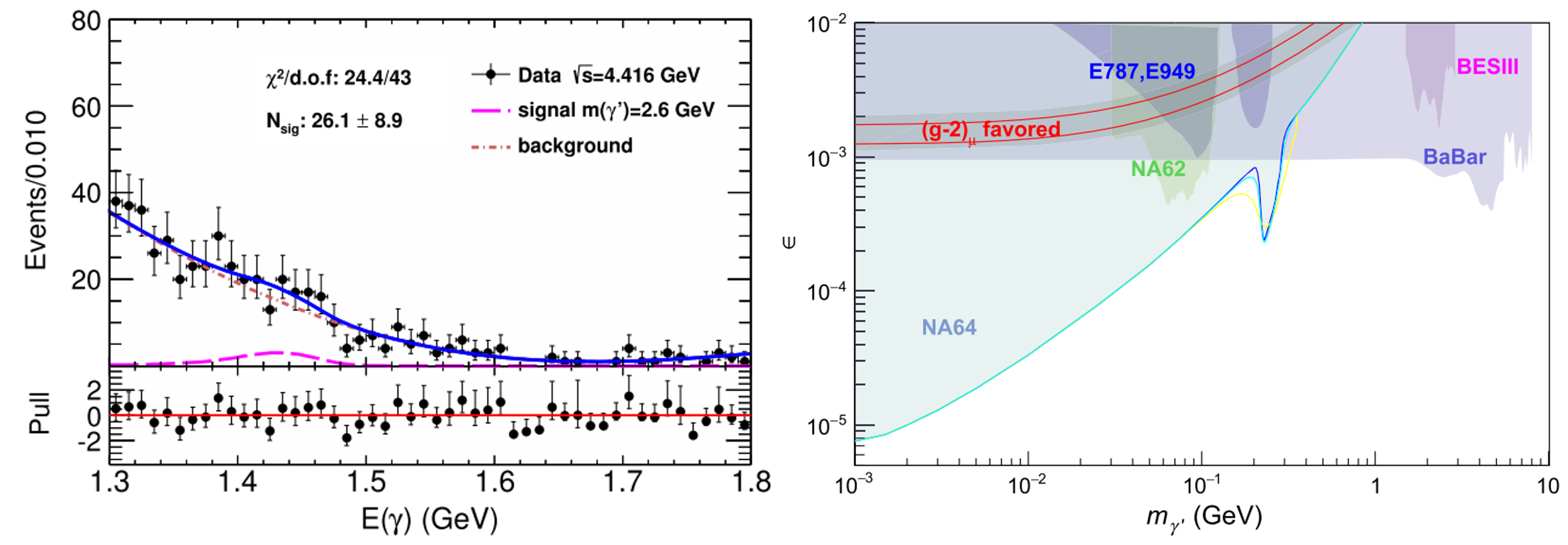}}

\caption{(Left) the representative plot of the fit to the $E(\gamma)$ spectra of data at $\sqrt{s} = 4.416$ GeV. (Right) the $90\%$ CL upper limits on $\epsilon$ as a function of $m_{\gamma'}$.   \label{dp}}
\end{figure*}

\noindent
{\bf\em\boldmath  Search for dark photon:}
The simplest scenerio of an Abelian $U(1)$ gauge field includes the possibility of a new type of force carrier called the dark photon ($\gamma'$). The dark photon is expected to couple to the SM particles through a kinetic mixing strength, defined as $\epsilon^2 = \alpha'/\alpha$, where $\alpha$ ($\alpha'$) is the fine structure constant in the SM (dark) sector~\cite{arkani}. The mass of the $\gamma'$ is expected to be a few GeV for satisfying the astrophysical constraints~\cite{dm} and the observed deviation in the muon anomalous magnetic moment up to the level of $4.2 \sigma$ between theory and experiment~\cite{anoml}. The exclusion limits on $\epsilon$  are constrained to be less than $10^{-3}$ by a series of experimental observations in both visible and invisible decays of dark photon~\cite{vindy}. BESIII has recently performed the search for invisible decays of  dark photon via initial-state-radiation (ISR) production of $e^+e^- \to \gamma \gamma'$ using 14.9 fb$^{-1}$ of $e^+e^-$ annihilation data taken at center-of-mass energies ($\sqrt{s}$) from 4.13 to 4.6 GeV~\cite{bes3dp}, where $\gamma$ is an ISR photon. The energy of this monochromatic ISR photon is calculated as, $E_{\gamma} = \frac{s - m_{\gamma'}^2 c^4}{2 \sqrt{s}}$, where $m_{\gamma'}$  is the mass of $\gamma'$. No  significant narrow resonance is observed in $E_{\rm ISR}$ spectrum of data (Fig.~\ref{dp} (left)). The mass dependent $90\%$ CL upper limits on $\epsilon$ for a dark photon coupling with an ordinary photon vary between $1.6 \times 10^{-3}$ and $5.7 \times 10^{-3}$ (Fig.~\ref{dp} (right)).

\begin{figure*}[htbp]
\centerline{\includegraphics[width=15.0cm]{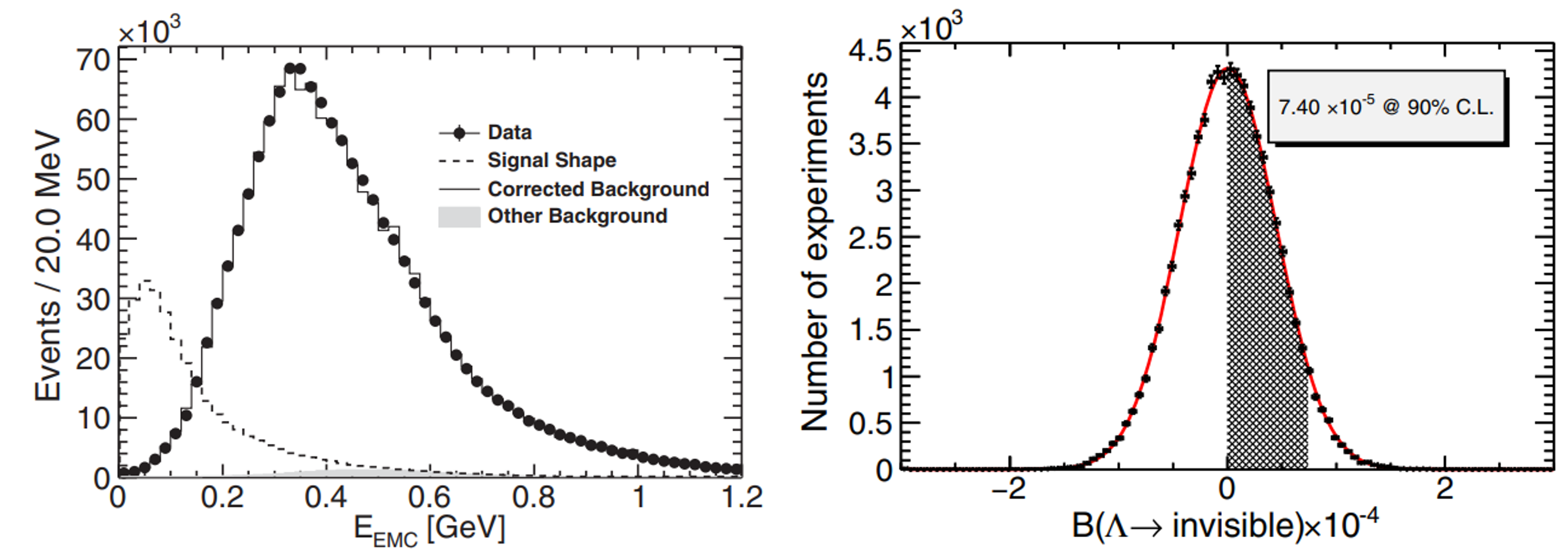}}

\caption{(Left) the $E_{\rm EMC}$ distribution of data (black dots with error bar), signal MC (dashed line) and background MC from $\Lambda \to n\pi^0$ decay (solid line).  (Right) the distribution of the estimated $\mathcal{B}(\Lambda \to {\rm invisible})$ obstained from pseudosamples.  \label{inv}}
\end{figure*}

\noindent
{\bf\em\boldmath  Invisible decays of \boldmath{$\Lambda$} baryon:}
The asymmetry between matter and antimatter in the Universe is a cause of baryon matter violation. The baryon matter and density and dark matter density are related as $\rho_{\rm DM} \approx 5.4 \rho_{\rm baryon}$. Thus, the dark matter may contribute to the invisible decays of baryon, such as $\Lambda \to {\rm invisible}$~\cite{barioDM}. The search for invisible decays of $\Lambda$ baryon has recently been performed with 10 billion of $J/\psi$ data through $J/\psi \to \Lambda \bar{\Lambda}$ for the first time using a double tag (DT) technique~\cite{bes3Linv}. In $J/\psi \to \Lambda \bar{\Lambda}$ decay, we first select single tag (ST) events in which $\Lambda$ baryon candidate is reconstructed with its dominant decay mode of $\Lambda \to p \pi$. Then the recoil side is used to infer the invisible decays of $\Lambda$ baryon. The signal events for invisible decays of $\Lambda$ baryon are extracted by performing a fit to the total energy deposited in the EMC ($E_{\rm EMC}$) distribution. Because of the background contribution from $\Lambda \to n \pi^0$, the $E_{\rm EMC}$ includes the following three contributions, $E_{\rm EMC} = E_{\rm EMC}^{\pi^0} + E_{\rm EMC}^n + E_{\rm EMC}^{\rm noise}$, where $E_{\rm EMC}^{\pi^0}$, $E_{\rm EMC}^n$ and $E_{\rm EMC}^{\rm noise}$ are the energy due to electromagnetic showers from $\pi^0$ decays, neutrons and shower unrelated to the events. The $E_{\rm EMC}$ distribution is expected to peak at zero for the signal-like events and deviated from zero for background from $\Lambda \to n \pi^0$, as seen in Fig.~\ref{inv} (left). No evidence of significant signal events is observed and $90\%$ CL upper limit on $\mathcal{B}(\Lambda \to {\rm invisible})$ is set to be less than $7.4 \times 10^{-5}$ shown in Fig.~\ref{inv} (right).

\begin{figure*}[htbp]
\centerline{\includegraphics[width=15.0cm]{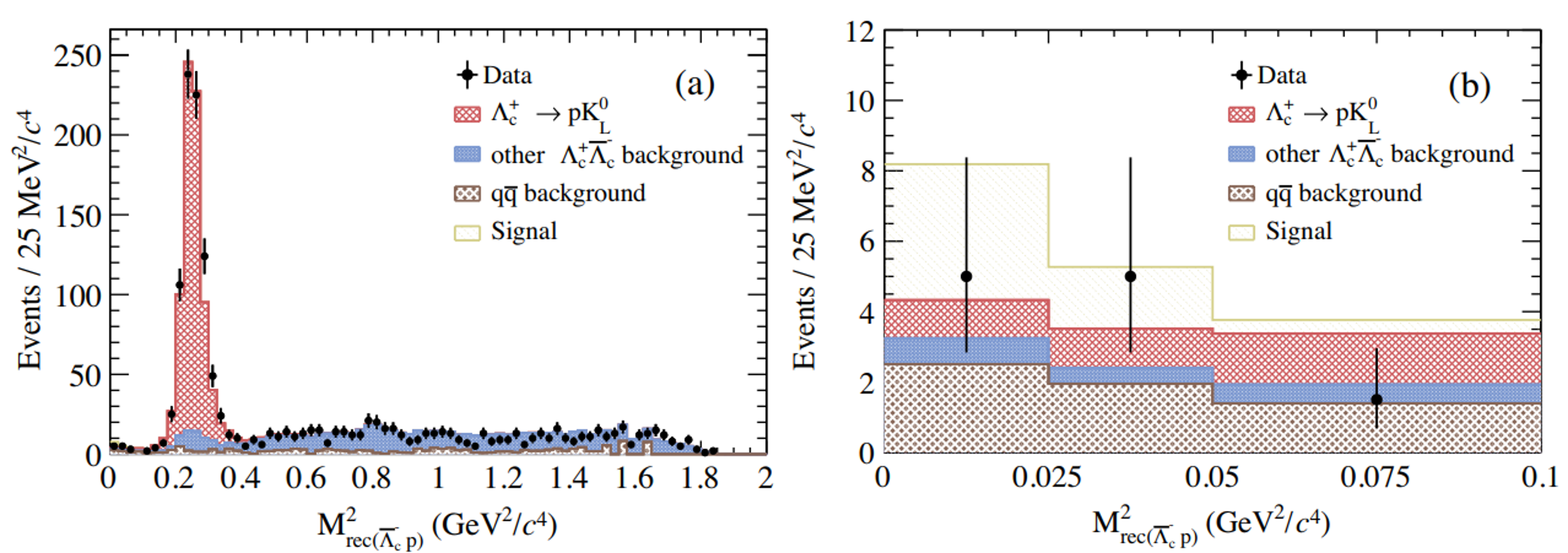}}
\caption{The $M^2_{{\rm rec}(\bar{\Lambda}^-_cp)}$ distribution of the accepted DT candidate events of data (black dots with error bar), signal MC (grey) and various background predictions shown by pattern coloured histograms in (left) full spectrum and (right) signal region.  \label{mrec}}
\end{figure*}

\noindent
{ \bf\em\boldmath   Search for massles dark photon:}
 A massless dark photon ($\gamma'$) is predicted by the spontaneous breaking of Abelian group $U(1)_D$~\cite{masslessdm, tandean}. The massless dark photon may also enhance the decay rate of FCNC processes in the charm-sector~\cite{glashow}. Such a massless dark photon can be accessible via two-body decay of $\Lambda_c \to p \gamma'$. BESIII has recently performed the search for massless dark photon via $e^+e^- \to \Lambda^+_c \bar{\Lambda}^-_c$, $\Lambda^+_c \to p \gamma'$ with 4.5 fb$^{-1}$ of data collected at center-of-mass energies between 4.6 and 4.699 GeV using a DT technique~\cite{bes3massless}. The data sample of $\bar{\Lambda}^-_c$ baryon, referred to as the ST sample, is reconstructed with its dominant hadronic decay modes and other $\Lambda^+_c$ baryon is allowed to decay via $\Lambda^+_c \to p \gamma'$. The DT events of the square of the recoil mass, $M^2_{{\rm rec}(\bar{\Lambda}^-_c p)}$, against the ST $\bar{\Lambda}_c^-$ and $p$ are utilized to infer the massless dark photon signal. The corresponding plot is shown in Fig.~\ref{mrec}.  No significant signal events for massless dark photon is observed. The $90\%$ CL upper limit on $\Lambda^+_c \to p \gamma')$ is set to be less than $8 \times 10^{-5}$.

\noindent
{\bf\em\boldmath  Summary:}
BESIII has performed the searches for dark matter and axion-like particles using the data samples collected at several energy points, including $J/\psi$ and $\psi(2S)$ resonances. No evidence of significant signal events is found in these scenarios, and set one of the most stringent exclusion limits. These exclusion limits are helpful to constrain a large fraction of the parameter spaces of the new physics models beyond the SM~\cite{barioDM, masslessdm, tandean, arkani, Essig, Merlo}. Many new results are expected to release soon with recently collected 20 fb$^{-1}$ of $\psi(3770)$ data.


\begin{thebibliography}{99}

\parskip=0pt plus 1pt minus 1pt

\bibitem{dm} O. Adriani {\it et al.}, Nature {\bf 458}, 607-609 (2009); M. Aguilar {\it et al.} (AMS Collaboration), Phys. Rev.
Lett. {\bf 110}, 141102 (2013); J. Chang {\it et al.}, Nature (London) {\bf 456}, 362 (2008); M. Ackermann {\it et al.}
(Fermi LAT Collaboration), Phys. Rev. Lett. {\bf 108}, 011103 (2012); M. Aguilar {\it et al.} (AMS
Collaboration), Phys. Rev. Lett. {\bf 110}, 141102 (2013). 
\bibitem{anoml} M. Pospelov, Phys. Rev. D {\bf 80} 095002 (2009).
 \bibitem{barioDM} J. Shelton and K. M. Zurek, Phys. Rev. D {\bf 82}, 123512 (2010)
\bibitem{masslessdm} M. Fabbrichesi, E. Gabrielli, and B. Mele, Phys. Rev. Lett. {\bf 119}, 031801 (2017).
\bibitem{tandean}  J. Y. Su and J. Tandean, Phys. Rev. D {\bf 101}, 035044 (2020).
 \bibitem{glashow} S. L. Glashow, J. Iliopoulos, and L. Maiani, Phys. Rev. D {\bf 2}, 1285 (1970).
\bibitem{arkani} N. Arkani-Hamed, D. P. Finkbeiner, T. R. Slatyer and N. Weiner, Phys. Rev. D {\bf 79}, 015014 (2009).
\bibitem{Essig} R. Essig {\it et al.}, Dark sectors and new, light, weakly-coupled particles, arXiv:1311.0029 (2013).
\bibitem{bes3} M. Abilikim {\it et al.} (BESIII Collaboration), Phys. Rev. D {\bf 93}, 052005 (2016).
\bibitem{Peccei}  R. D. Peccei and H. R. Quinn, Phys. Rev. Lett. {\bf 38}, 1440 (1977); Phys. Rev. D {\bf 16}, 1791 (1977).
\bibitem{Weinberg} S. Weinberg, Phys. Rev. Lett. 40, 223 (1978); F. Wilczek, Phys. Rev. Lett. 40, 279 (1978).
\bibitem{Graham} P. W. Graham, D. E. Kaplan and S. Rajendran, Phys. Rev. Lett. {\bf 115}, 221801 (2015). 
\bibitem{higgs} G. C. Branco et al., Phys. Rept. {\bf 516}, 1 (2012).
\bibitem{string} A. Ringwald, J. Phys. Conf. Ser. {\bf 485}, 012013 (2014).
\bibitem{Merlo} L. Merlo, F. Pobbe et al., J. High Energy Phys. 06, 091 (2019). 
\bibitem{Masso} E. Masso and R. Toldra, Phys. Rev. D 52, 1755 (1955).
\bibitem{opal} S. Knapen {\it et al.}, Phys. Rev. Lett. {\bf 118}, 171801 (2017).
\bibitem{belle2} F. Abudinen {\it et al.} [Belle II Collaboration], Phys. Rev. Lett. 125, 161806 (2020).
\bibitem{bes3alp}  M. Ablikim et al. [BESIII Collaboration], Phys. Lett. B {\bf 44}, 137678 (2023).
\bibitem{nJps}  M. Ablikim et al. [BESIII Collaboration], Chin. Phys. C {\bf 46}, 074001 (2022).
\bibitem{Maniatis}  M. Maniatis, Int. J. Mod. Phys. A {\bf 25}, 3505-3602 (2010).
\bibitem{Wilczek}  F. Wilczek, Phys. Rev. Lett. {\bf 39}, 1304 (1977).
\bibitem{Dermisek} R. Dermisek, J. F. Gunion and B. McElrath, Phys. Rev. D {\bf 76}, 051105 (2007).
\bibitem{prevbes3} M. Abilikim et al. (BESIII Collaboration), Phys. Rev. D {\bf 93}, 052005 (2016).
\bibitem{baber} J. P. Lees et al. (BaBar Collaboration), Phys. Rev. D {\bf 87}, 031102(R) (2013).
\bibitem{bes3higgs} M. Abilikim et al. (BESIII Collaboration), Phys. Rev. D {\bf 105}, 012008 (2022).
\bibitem{vindy} V. Prasad, Dark matter/new physics searches at BESIII, POS (ALPS2019) 030.
\bibitem{bes3dp} M. Abilikim et al. (BESIII Collaboration), Phys. Lett. B {\bf 839}, 137785 (2023).
\bibitem{bes3Linv} M. Abilikim et al. (BESIII Collaboration), Phys. Rev. D {\bf 105}, L071101 (2022).
\bibitem{bes3massless} M. Abilikim et al. (BESIII Collaboration), Phys. Rev. D {\bf 106}, 072008 (2022).

\end{thebibliography}
\end{document}